\documentclass{PoS}

\title{Pauli--Villars regularization of field theories on the light front}

\ShortTitle{Pauli--Villars regularization of field theories}

\author{\speaker{John HILLER}\\
        University of Minnesota-Duluth, USA\\
        E-mail: \email{jhiller@d.umn.edu}}

\abstract{%
Four-dimensional quantum field theories generally require
regularization to be well defined.  This can be done in
various ways, but here we focus on Pauli--Villars (PV)
regularization and apply it to nonperturbative calculations
of bound states.  The philosophy is to introduce enough
PV fields to the Lagrangian to regulate the theory
perturbatively, including preservation of symmetries,
and assume that this is sufficient for the nonperturbative
case.  The numerical methods usually necessary for
nonperturbative bound-state problems are then applied
to a finite theory that has the original symmetries.  The
bound-state problem is formulated as a mass eigenvalue
problem in terms of the light-front Hamiltonian.  
Applications to quantum electrodynamics
are discussed.
}

\FullConference{Light Cone 2010 - LC2010\\
		June 14-18, 2010\\
		Valencia, Spain}

\def\bea{\begin{eqnarray}}
\def\eea{\end{eqnarray}}
\def\be{\begin{equation}}
\def\ee{\end{equation}}
\newcommand{\ub}[1]{\underline{#1}}

\begin{document}

\section{Introduction}
\label{sec:Introduction}

In order to solve a ($3+1$)-dimensional theory,
the theory must be regulated in some way.  In
doing so, one should attempt to preserve as many symmetries as possible.
We do this by adding enough Pauli--Villars (PV)~\cite{PauliVillars} 
fields to regulate perturbation theory and assume that the nonperturbative
eigenproblem is also regulated.  Numerical methods are then applied to
a finite theory, just as was the case for (1+1)-dimensional 
superrenormalizable theories~\cite{DLCQreview}.  From the Hamiltonian eigenproblem,
we can compute wave functions as coefficients in Fock-state expansions,
and from these, compute observables.

As a test of the approach, we consider QED.  This is not meant
to compete with perturbation theory; the numerical errors of
the nonpertubative calculation are too large to resolve high-order
contributions that perturbation theory computes directly.  However,
the method is intended for strong-coupling theories where perturbation
theory is ineffective.

There have been a number of applications of the method.  The
first being an exploration in terms of a soluble, heavy-source
model~\cite{bhm1,bhm2}.  The dressed-fermion state in Yukawa theory
has been studied extensively~\cite{YukawaDLCQ,YukawaOneBoson,YukawaTwoBoson},
as have exact solutions in the limit of equal PV masses~\cite{ExactSolns}.
Applications to gauge theories have been primarily to the dressed-electron
state in QED~\cite{OnePhotonQED,ChiralLimit,Thesis,SecDep,TwoPhotonQED}, but
also to the photon eigenstate~\cite{VacPol}.  A scheme has been proposed
for QCD~\cite{Paston}.

In order to have a well-defined Fock-state expansion, the
theories are quantized on the light front~\cite{Dirac,DLCQreview}.
We define light-cone coordinates
of time, $x^+=t+z$, and space, $\ub{x}=(x^-,\vec{x}_\perp)$,
with   $x^-\equiv t-z$ and $\vec{x}_\perp=(x,y)$.  The light-cone
energy is $p^-=E-p_z$ and momentum, $\ub{p}=(p^+,\vec{p}_\perp)$,
with $ p^+\equiv E+p_z$ and $\vec{p}_\perp=(p_x,p_y)$.  These
lead to the mass-shell condition, $p^2=m^2$,
in the form $p^-=\frac{m^2+p_\perp^2}{p^+}$.

We work with the standard parameterization, where the
bare parameters of the Lagrangian are fixed by fits
to physical constraints.  There is the alternative
of sector-dependent parameterization, where
the bare parameters of the Lagrangian are allowed to
depend on the Fock sector(s) on which the operators
act.  This was originally proposed by Perry, Harindranath,
and Wilson~\cite{SectorDependent} and applied to QED
by Hiller and Brodsky~\cite{hb}.  More recent work
with this approach has been by Karmanov, Mathiot, and
Smirnov~\cite{Karmanov,Karmanov2010}.
Other nonperturbative approaches include lattice 
theory~\cite{Lattice,TransLattice},
Dyson--Schwinger equations~\cite{DSE},
effective-particle representations~\cite{Glazek},
and basis-function methods~\cite{Vary}.

The remainder of this paper contains a brief description
of light-front QED in Lorentz gauge, followed by discussion
of the eigenproblem for the electron state
in lowest truncation.
The anomalous magnetic moment is calculated.  We also compare
the results of the sector-dependent approach, and then
conclude with a brief summary.

\section{Light-front QED in Lorentz gauge}
\label{sec:LightFrontQED}

The Lorentz-gauge QED Lagrangian, regulated by two PV fermion 
flavors and two PV photon flavors, is
\bea \label{eq:Lagrangian}
{\cal L} &= & \sum_{i=0}^2 (-1)^i \left[-\frac14 F_i^{\mu \nu} F_{i,\mu \nu} 
         +\frac12 \mu_i^2 A_i^\mu A_{i\mu} 
         -\frac{1}{2} \left(\partial^\mu A_{i\mu}\right)^2\right] \\
  && + \sum_{i=0}^2 (-1)^i \bar{\psi_i} (i \gamma^\mu \partial_\mu - m_i) \psi_i 
  - e_0 \bar{\psi}\gamma^\mu \psi A_\mu ,  \nonumber
\eea
where
\begin{equation} \label{eq:NullFields}
  \psi =  \sum_{i=0}^2 \sqrt{\beta_i} \psi_i, \;\;
  A_\mu  = \sum_{i=0}^2 \sqrt{\xi_i}A_{i\mu}, \;\;
  F_{i\mu \nu} = \partial_\mu A_{i\nu}-\partial_\nu A_{i\mu} .
\end{equation}
A subscript of $i=0$ indicates a physical field, and $i=1$ or 2 a PV field.
The $i=1$ fields are chosen to have negative norm.  The mass of the
bare photon $\mu_0$ is zero.

The constants $\beta_i$ and $\xi_i$ control the coupling strengths
of the various fields.  We require that $\beta_0=1$ and
$\xi_0=1$ and require the constraints
$\sum_{i=0}^2(-1)^i\beta_i=0$ and $\sum_{i=0}^2(-1)^i\xi_i=0$.
These guarantee the regularization and
that the combinations $\psi$ and $A_\mu$
in (\ref{eq:NullFields}) have zero norm.  A third pair of
constraints comes from requiring that the photon eigenstate
have zero mass~\cite{VacPol} and that the mass of the electron eigenstate
becomes zero when $m_0$ is set to zero~\cite{ChiralLimit}.

The nondynamical fermion fields satisfy the following
constraints ($i=0$,1,2):
\be  \label{eq:Psii-Constraint}
i(-1)^i\partial_-\psi_{i-}+e_0 A_-\sqrt{\beta_i}\psi_- 
  =i\gamma^0\vec\gamma^\perp\cdot
     \left[(-1)^i\vec\partial_\perp \psi_{i+}
         -ie_0 \sqrt{\beta_i}\vec A_\perp\psi_+\right] 
      -(-1)^i m_i \gamma^0\psi_{i+} . 
\ee
From these we obtain a constraint on the null combination
\be
i\partial_-\psi_-
  =(i\gamma^0\vec\gamma^\perp)\cdot
     \vec\partial_\perp \psi_+
      - \gamma^0\sum_i m_i \sqrt{\beta_i}\psi_{i+}.
\ee
The terms containing the photon field cancel because $\sum_i(-1)^i\beta_i=0$;
therefore, light-cone gauge is not necessary.
The nondynamical field $\psi_-$ can then be constructed from a sum
of $\psi_{i-}$ that satisfy the free-fermion constraint.

The mode expansion for the full Fermi field is
\be
\psi_i=\frac{1}{\sqrt{16\pi^3}}\sum_s\int \frac{d\ub{k}}{\sqrt{k^+}} 
  \left[b_{is}(\ub{k})e^{-i\ub{k}\cdot\ub{x}}u_{is}(\ub{k})
        +d_{i,-s}^\dagger(\ub{k})e^{i\ub{k}\cdot\ub{x}}v_{is}(\ub{k})
        \right].
\ee
The spinors are defined in \cite{LepageBrodsky},
and the nonzero anticommutators are \\
$\{b_{is}(\ub{k}),b_{i's'}^\dagger(\ub{k}')\}
   =(-1)^i\delta_{ii'}\delta_{ss'}\delta(\ub{k}-\ub{k}')$ and
$\{d_{is}(\ub{k}),d_{i's'}^\dagger(\ub{k}')\}
   =(-1)^i\delta_{ii'}\delta_{ss'}\delta(\ub{k}-\ub{k}')$.

The mode expansion for the $i$th photon flavor is
\be
A_{i\mu}=\frac{1}{\sqrt{16\pi^3}}\int \frac{d\ub{k}}{\sqrt{k^+}}
  \left[a_{i\mu}(\ub{k})e^{-i\ub{k}\cdot\ub{x}}
        +a_{i\mu}^\dagger(\ub{k})e^{i\ub{k}\cdot\ub{x}}\right] ,
\ee
with the commutator
${[}a_{i\mu}(\ub{k}),a_{i'\nu}^\dagger(\ub{k}')]
   =(-1)^i\delta_{ii'}\epsilon^\mu\delta_{\mu\nu}\delta(\ub{k}-\ub{k}')$.
The metric signature is chosen to be $\epsilon^\mu = (-1,1,1,1)$, for
Gupta--Bleuler quantization~\cite{GuptaBleuler}.
Because we do not use light-cone gauge, there is no constraint
on $A_+=A^-$, and, consequently, there will be no instantaneous
photon interaction term~\cite{DLCQreview} in the Hamiltonian.

We can now construct the light-front Hamiltonian ${\cal P}^-$
from spinor matrix elements~\cite{VacPol}:
\bea \label{eq:QEDP-}
{\cal P}^-&=&
   \sum_{i,s}\int d\ub{p}
      \frac{m_i^2+p_\perp^2}{p^+}(-1)^i
          b_{i,s}^\dagger(\ub{p}) b_{i,s}(\ub{p}) 
    +\sum_{i,s}\int d\ub{p}
      \frac{m_i^2+p_\perp^2}{p^+}(-1)^i
          d_{i,s}^\dagger(\ub{p}) d_{i,s}(\ub{p}) \nonumber \\
   && +\sum_{l,\mu}\int d\ub{k}
          \frac{\mu_l^2+k_\perp^2}{k^+}(-1)^l\epsilon^\mu
             a_{l\mu}^\dagger(\ub{k}) a_{l\mu}(\ub{k})
          \\
   && +\sum_{i,j,l,s,\mu}\sqrt{\beta_i\beta_j\xi_l}\int d\ub{p} d\ub{q}\left\{
      b_{i,s}^\dagger(\ub{p}) \left[ b_{j,s}(\ub{q})
       V^\mu_{ij,2s}(\ub{p},\ub{q})\right.\right.\nonumber \\
      &&\left.\left.\rule{1.75in}{0in}
+ b_{j,-s}(\ub{q})
      U^\mu_{ij,-2s}(\ub{p},\ub{q})\right] 
            a_{l\mu}^\dagger(\ub{q}-\ub{p})  \right. \nonumber \\
&& +     b_{i,s}^\dagger(\ub{p}) \left[ d_{j,s}^\dagger(\ub{q})
       \bar V^\mu_{ij,2s}(\ub{p},\ub{q}) 
+ d_{j,-s}^\dagger(\ub{q})
      \bar U^\mu_{ij,-2s}(\ub{p},\ub{q})\right] 
            a_{l\mu}(\ub{q}+\ub{p})
                      \nonumber \\
&& -   \left.   d_{i,s}^\dagger(\ub{p}) \left[ d_{j,s}(\ub{q})
       \tilde V^\mu_{ij,2s}(\ub{p},\ub{q}) 
+ d_{j,-s}(\ub{q})
      \tilde U^\mu_{ij,-2s}(\ub{p},\ub{q})\right] 
            a_{l\mu}^\dagger(\ub{q}-\ub{p})
                    + H.c.\right\} .  \nonumber
\eea
The vertex functions can be found in \cite{VacPol}.
The Hamiltonian does not contain any instantaneous fermion terms~\cite{DLCQreview}.
They cancel between physical and PV contributions because they are independent
of the fermion mass and proportional to $(-1)^i\beta_i$ for the $i$th flavor.
The sum over flavors then yields $\sum_i(-1)^i\beta_i=0$.  This is
independent of the gauge choice.

\section{Electron eigenstate}
\label{sec:Electron}

In the one-electron/one-photon truncation,
the Fock-state expansion of the electron eigenstate, for total
$J_z=\pm\frac12$, is
\be \label{eq:FockExpansion}
|\psi^\pm(\ub{P})\rangle=\sum_i z_i b_{i\pm}^\dagger(\ub{P})|0\rangle
  +\sum_{ijs\mu}\int d\ub{k} C_{ijs}^{\mu\pm}(\ub{k})b_{is}^\dagger(\ub{P}-\ub{k})
                                       a_{j\mu}^\dagger(\ub{k})|0\rangle.
\ee
It is normalized according to 
$\langle\psi^{\sigma'}(\ub{P}')|\psi^\sigma(\ub{P})\rangle
                =\delta(\ub{P}'-\ub{P})\delta_{\sigma'\sigma}$.
The second PV fermion flavor ($i=2$) plays no role in this
sector and can be removed.  In order to have a positive
norm, all physical quantities are computed from a
projected state~\cite{OnePhotonQED}:
\bea
|\psi_{\rm phys}^\pm(\ub{P})\rangle&=&\sum_i (-1)^i z_i
                                          b_{0\pm}^\dagger(\ub{P})|0\rangle
+\sum_{s\mu}\int d\ub{k} \sum_{i=0}^1\sum_{j=0,2}\sqrt{\xi_j} \\
&&\times
        \sum_{k=j/2}^{j/2+1} \frac{(-1)^{i+k}}{\sqrt{\xi_k}}
                        C_{iks}^{\mu\pm}(\ub{k})
               b_{0s}^\dagger(\ub{P}-\ub{k})
                                       a_{j\mu}^\dagger(\ub{k})|0\rangle \nonumber.
\eea

To be an eigenstate of the light-cone Hamiltonian,
the wave functions must satisfy the following coupled
equations, with $y=k^+/P^+$:
\bea \label{eq:firstcoupledequation}
[M^2-m_i^2]z_i & = & \int (P^+)^2 dy d^2k_\perp 
     \sum_{j,l,\mu}\sqrt{\xi_l}(-1)^{j+l}\epsilon^\mu \\
&& \times
  \left[V_{ji\pm}^{\mu*}(\ub{P}-\ub{k},\ub{P})C^{\mu\pm}_{jl\pm}(\ub{k}) 
        +U_{ji\pm}^{\mu*}(\ub{P}-\ub{k},\ub{P}) C^{\mu\pm}_{jl\mp}(\ub{k})\right] , \nonumber
\eea
and
\be \label{eq:TwoBodyEqns}
\left[M^2 - \frac{m_j^2 + k_\perp^2}{1-y} - \frac{\mu_l^2 + k_\perp^2}{y}\right]
  C^{\mu\pm}_{jl\pm}(\ub{k})
    =\sqrt{\xi_l}\sum_{i'} (-1)^{i'} z_{i'} P^+ V_{ji'\pm}^\mu(\ub{P}-\ub{k},\ub{P}),
\ee
\be \label{eq:TwoBodyEqns2}
\left[M^2 - \frac{m_j^2 + k_\perp^2}{1-y} - \frac{\mu_l^2 + k_\perp^2}{y}\right]
C^{\mu\pm}_{jl\mp}(\ub{k}) 
    = \sqrt{\xi_l}\sum_{i'} (-1)^{i'} z_{i'} P^+ U_{ji'\pm}^\mu(\ub{P}-\ub{k},\ub{P}).
\ee
An index of $i$ corresponds to the 
one-electron sector and $j$ to the one-electron/one-photon sector.  
Therefore, in the sector-dependent approach, a mass $m_i$ in a vertex
function is assigned the bare mass, and $m_j$ is the physical
mass.  In the standard parameterization, all are bare masses.

The coupled equations can be solved analytically~\cite{OnePhotonQED}. 
The wave functions $C_{ils}^{\mu\pm}$ are
\begin{eqnarray} \label{eq:wavefn1}
C^{\mu\pm}_{il\pm}(\ub{k}) &=& \sqrt{\xi_l}
  \frac{\sum_j (-1)^j z_j P^+ V_{ij\pm}^\mu(\ub{P}-\ub{k},\ub{P})}
    {M^2 - \frac{m_i^2 + k_\perp^2}{1-y} - \frac{\mu_l^2 + k_\perp^2}{y}} , \\
\label{eq:wavefn2}
C^{\mu\pm}_{il\mp}(\ub{k}) &=& \sqrt{\xi_l}
\frac{\sum_j (-1)^j z_j P^+ U_{ij\pm}^\mu(\ub{P}-\ub{k},\ub{P})}
     {M^2 - \frac{m_i^2 + k_\perp^2}{1-y} - \frac{\mu_l^2 + k_\perp^2}{y}} ,
\end{eqnarray}
and the amplitudes satisfy
\be \label{eq:FeynEigen}
(M^2-m_i^2)z_i =
      2e_0^2\sum_{i'} (-1)^{i'}z_{i'}\left[\bar{J}+m_im_{i'} \bar{I}_0
  -2(m_i+m_{i'}) \bar{I}_1 \right],
\ee
with~\cite{OnePhotonQED}
\begin{eqnarray} \label{eq:In}
\bar{I}_n(M^2)&=&\int\frac{dy dk_\perp^2}{16\pi^2}
   \sum_{jl}\frac{(-1)^{j+l}\xi_l}{M^2-\frac{m_j^2+k_\perp^2}{1-y}
                                   -\frac{\mu_l^2+k_\perp^2}{y}}
   \frac{m_j^n}{y(1-y)^n}\,, \\
\bar{J}(M^2)&=&\int\frac{dy dk_\perp^2}{16\pi^2}  \label{eq:J}
   \sum_{jl}\frac{(-1)^{j+l}\xi_l}{M^2-\frac{m_j^2+k_\perp^2}{1-y}
                                   -\frac{\mu_l^2+k_\perp^2}{y}}
   \frac{m_j^2+k_\perp^2}{y(1-y)^2} .
\end{eqnarray}
There is also the identity~\cite{ChiralLimit} $\bar{J}=M^2 \bar{I}_0$.

The analytic solution is~\cite{OnePhotonQED}
\be \label{eq:OneBosonEigenvalueProb}
\alpha_{0\pm}=\frac{(M\pm m_0)(M\pm m_1)}{8\pi (m_1-m_0)(2 \bar{I}_1\pm M\bar{I}_0)} , \;\;
z_1=\frac{M \pm m_0}{M \pm m_1}z_0 .
\ee
A graphical solution is given in \cite{ChiralLimit}.  In general,
the lower sign yields the physical answer, because $m_0$ then
becomes the physical mass $M=m_e$ at zero coupling.
The coupling coefficient $\xi_2$ is fixed by requiring that
$M=0$ when $m_0=0$.  In this truncation,
we can safely take the 
$m_1\rightarrow\infty$ limit, where $z_1=0$, $m_1z_1\rightarrow\pm(M-m_0)z_0$,
and
\be  \label{eq:alpha0}
\alpha_{0\pm}=\pm\frac{M(M\pm m_0)}{8\pi (2 \bar{I}_1\pm M\bar{I}_0)},
\ee
and the second PV photon flavor can be discarded.
In the sector-dependent approach, $\bar I_1$ and $\bar I_0$ are independent
of $m_0$, and the solution for $\alpha_0$ 
can be rearranged as an explicit expression for $m_0$
\be
m_0=\mp M + 8\pi\frac{\alpha_{0\pm}}{M}(2 \bar{I}_1\pm M\bar{I}_0).
\ee

The anomalous magnetic moment is computed from the spin-flip matrix element
of the electromagnetic current $J^+$~\cite{BrodskyDrell}.
In the one-photon truncation and in the limit where the PV electron mass 
$m_1$ is infinite, the expression for the anomalous moment is
\bea
a_e&=&\frac{\alpha_0}{\pi}m_e^2z_0^2\int y^2 (1-y) dy dk_\perp^2 \\
  &&  \times
  \left(\sum_{k=0}^1 \frac{(-1)^k}{ym_0^2+(1-y)\mu_k^2+k_\perp^2-m_e^2y(1-y)}\right)^2 .
    \nonumber
\eea
For the sector-dependent parameterization, the product $\alpha_0 z_0^2$ is just $\alpha$,
and the bare mass $m_0$ in the denominator is replaced by the physical mass $m_e$.
Plots of the anomalous moment as a function of the regulator mass $\mu_1$ can
be found in \cite{SecDep}.

\section{Sector-dependent and standard parameterizations}
\label{sec:SectorDependent}

Even though we do not include the vacuum-polarization 
contribution to the dressed-electron state, the
sector-dependent bare coupling is not equal to the
physical coupling.  Instead, they are related by
$e_0=e/z_0$~\cite{hb}, where $z_0$ is the amplitude for the
bare-electron Fock state computed without projection onto
the physical subspace.  In general, the bare coupling
would be $e_0=Z_1e/\sqrt{Z_{2i}Z_{2f}Z_3}$; this includes
the truncation effect that splits the usual $Z_2$ into
a product of different $\sqrt{Z_2}$ from each fermion
leg~\cite{OSUQED}.  With no fermion-antifermion loop,
we have $Z_3=1$, and without a second photon, there is  
no vertex correction and $Z_1=1$.
Also, only the fermion leg with no photon spectator
will be corrected by $\sqrt{Z_2}$;
therefore, we find $\sqrt{Z_{2i}Z_{2f}}=z_0$.

In the sector-dependent approach, the
bare-electron amplitude without projection is determined,
in the infinite-$m_1$ limit, by the normalization
$1=z_0^2+e_0^2z_0^2\tilde J_2$, with
\be \label{eq:tildeJ2}
\tilde J_2=\frac{1}{8\pi^2}\int y\, dy dk_\perp^2
  \sum_{k=0}^1(-1)^{k}
  \frac{(y^2+2y-2)m_e^2+k_\perp^2}
       {[k_\perp^2+(1-y)\mu_{k}^2+y^2m_e^2]^2}.
\ee
Replacing $e_0$ by $e/z_0$, we can solve for $z_0$ as
$z_0=\sqrt{1-e^2\tilde J_2}$ and find
$e_0=e/\sqrt{1-e^2\tilde J_2}$.
For large $\mu_1$, one finds
$\tilde J_2\simeq \frac{1}{8\pi^2}
\left(\ln\frac{\mu_1\mu_0^2}{m_e^3}+\frac98\right)$.
Thus, $e_0$ can become imaginary and Fock-sector
probabilities range outside $[0,1]$
due to IR and UV divergences, and consistency
then imposes limits on $\mu_0$ and $\mu_1$~\cite{SecDep},
as confirmed for Yukawa theory in \cite{Karmanov2010}.

In the standard parameterization, the bare amplitude
is determined by $1=z_0^2+e^2z_0^2 J_2$, with
\bea \label{eq:J2}
J_2&=&\frac{1}{8\pi^2}\int y\, dy dk_\perp^2
   [m_0^2-4m_0m_e(1-y)+m_e^2(1-y)^2+k_\perp^2] \\
 && \times \left(\sum_{k=0}^1(-1)^k
  \frac{1}
       {[k_\perp^2+(1-y)\mu_{k}^2+ym_0^2-y(1-y)m_e^2]}\right)^2 .  \nonumber
\eea
Thus the bare amplitude is $z_0=1/\sqrt{1+e^2J_2}$,
which is driven to zero as $\mu_1\rightarrow\infty$ and causes
most expectation values also to go to zero.  Therefore,
in this case there is a limit on $\mu_1$, but $\mu_0$ can be zero.

The anomalous moment in the sector-dependent case is
\bea \label{eq:secdepae}
a_e&=&\frac{\alpha}{\pi}m_e^2\int y^2 (1-y) dy dk_\perp^2 \\
  &&  \sum_{k=0}^1 (-1)^k
    \left( \frac{1}{ym_e^2+(1-y)\mu_k^2+k_\perp^2-m_e^2y(1-y)}\right)^2
    \nonumber
\eea
In the $\mu_1\rightarrow\infty$,
$\mu_0\rightarrow0$ limit, this becomes
exactly the Schwinger result
\be \label{eq:Schwinger}
a_e=\frac{\alpha}{\pi}m_e^2\int
     \frac{ dy dq_\perp^2/(1-y)}{\left[\frac{m_e^2+q_\perp^2}{1-y}
     +\frac{q_\perp^2}{y}-m_e^2\right]^2} = \frac{\alpha}{2\pi}. 
\ee
However, this limit cannot be taken without making the underlying
theory inconsistent.

\section{Summary}
\label{sec:Summary}

With use of PV regularization, one can formulate
and solve nonperturbative problems in field theories.
It is important to maintain symmetries, which can
be done with additional PV fields, such as
those introduced to maintain a zero photon mass
and the chiral symmetry of the massless-electron limit.
It is best to regulate before applying numerical methods, to clearly
separate limits of regulators from those of numerical convergence.
The PV fields do add to the numerical load but also reduce it,
by eliminating instantaneous fermion and instantaneous photon
interactions.

For both the standard parameterization and the
sector-dependent parameterization,
truncation of the Fock space results in uncancelled
divergences, which require that not all PV masses
be taken to infinity; however, meaningful results
can be extracted at finite PV masses.  For the
sector-dependent approach, this is complicated
by infrared divergences~\cite{SecDep}.

As discussed elsewhere~\cite{TwoPhotonQED}, these methods
have been extended to a truncation that includes
two photons.  The next step is to also include an
electron-positron-pair contribution to the dressed-electron
state and study charge renormalization
as well as current covariance.  This truncation
will then include all contributions
of order $\alpha^2$ to the anomalous moment.
A calculation at large $\alpha$, where numerical
errors in the order-$\alpha$ contribution would be small
compared to the $\alpha^2$ contributions,
could be compared with higher-order perturbation theory.
It would also be interesting to consider the dressed
electron in a magnetic field and extract its
induced magnetic moment.  In addition, as a precursor
to consideration of mesons in QCD, one can compute
two-fermion bound states in Yukawa theory and QED.

\acknowledgments
The work reported here was done in collaboration with 
S.S. Chabysheva
and supported in part by the US Department of Energy
and the Minnesota Supercomputing Institute.

\end{document}